\documentclass[twocolumn,amsmath,amssymb, nobibnotes, aps, prl,showkeys]{revtex4}
\usepackage{graphicx}
\usepackage{dcolumn}
\usepackage{bm}
\usepackage{docs}
\usepackage{booktabs}%
\usepackage{algorithm}%
\usepackage{algorithmicx}%
\usepackage{listings}%
\usepackage{footnote}
\usepackage{tabularx}

\expandafter\ifx\csname package@font\endcsname\relax\else
 \expandafter\expandafter
 \expandafter\usepackage
 \expandafter\expandafter
 \expandafter{\csname package@font\endcsname}%
\fi
\newcommand{\be}{\begin{equation}}
\newcommand{\ee}{\end{equation}}
\newcommand{\bea}{\begin{eqnarray}}
\newcommand{\eea}{\end{eqnarray}}

\begin{document}
\title{Ordinary Stars as Potential TeV Cosmic-Ray Accelerators}

\author{Prabir Banik$^{1}$ \thanks{Email address: pbanik74@yahoo.com}, Arunava Bhadra$^{1}$\email{Email address: aru\_bhadra@yahoo.com} and Sanjay K. Ghosh$^{2}$ \thanks{Email address: sanjay@jcbose.ac.in}}
\affiliation{$^{1}$High Energy $\&$ Cosmic Ray Research Centre, University of North Bengal, Siliguri, West Bengal, India 734013. \\
$^{2}$Department of Physics, $\&$ Center for Astroparticle Physics $\&$ Space Science,Bose Institute, EN-80, Sector-5, Bidhan Nagar,  Kolkata-700091, India.}

\begin{abstract}
Recent observations of cosmic rays increasingly point to the existence of nearby sources $-$ so-called ``local tevatrons", capable of accelerating particles to TeV energies. In this study, we examine the potential of a typical main-sequence star, represented by the Sun, to act as a source of TeV cosmic rays (CRs). We focus on identifying plausible mechanisms through which a quiescent star can accelerate charged particles to relativistic energies. We show that shock-drift acceleration processes operating within the chromospheres of the Sun and similar stars can accelerate particles to energies reaching the TeV scale. Additionally, we provide quantitative estimates of both the maximum achievable particle energies, spectral index of energy spectrum and the resulting cosmic-ray fluxes that such stellar environments could realistically produce. Our results indicate that ordinary stars could potentially contribute to the fine structure observed in the cosmic-ray spectrum at TeV energies and may help explain the local excess of TeV-scale electrons and positrons detected by H.E.S.S. and other observatories.  
\end{abstract}

\maketitle

\section{Introduction}\label{sec1}
The predominant paradigm in cosmic ray (CR) astrophysics posits that the majority of observed primary CRs are accelerated, potentially up to energies of approximately 3 PeV, corresponding to the so-called ``knee" in the CR spectrum, at supernova remnants (SNRs) within our galaxy via diffusive shock acceleration (DSA) \cite{Berezinsky90,Blasi13}. This framework is supported by energy budget considerations, as SNRs are capable of supplying the requisite energy to sustain the observed CR energy density in the Milky Way. Under the DSA mechanism, the resultant spectrum of accelerated particles is expected to follow a featureless power-law distribution.

Recent high-precision measurements, however, increasingly challenge the sufficiency of this conventional model. Observations have revealed notable deviations from a simple power-law behavior in the CR proton spectrum, specifically a spectral hardening at energies of approximately 200 GeV \cite{Adriani11,Aguilar15,An19,Adriani22} and a softening around 10 TeV or a slightly higher energy \cite{An19,Adriani22,Choi22,Grebenyuk19}. These features present a significant theoretical challenge, as the long residence times of CRs in the Galaxy should lead to smooth and continuous energy spectra. To account for the observed fine structure at TeV energies, various studies have proposed the contribution of local, nearby sources \cite{Zatsepin06,Gaisser13,Tomassetti15}.

Additionally, the spectra of CR electrons and positrons have been measured \cite{Aguilar14,Adriani13,Ackerman12} across a wide energy range, from a few GeV to several tens of TeV, by modern space-based experiments. At energies above 100 GeV, electrons and positrons suffer significant radiative energy losses due to synchrotron radiation and inverse-Compton scattering while propagating through the interstellar medium \cite{Aharonian24}. These loss mechanisms constrain their cooling times and hence their effective propagation distances to typically less than a kiloparsec, within a diffusion-dominated transport regime. Consequently, high-energy CR electrons and positrons detected near Earth are believed to originate from one or more local primary accelerators \cite{Mauro14,Recchia19}.

Although a main-sequence star such as the Sun also generates CRs during episodes of flaring activity, the energies attained by these solar-accelerated particles are generally limited to a few GeV \cite{Biswas00,Dorman21}. However, recent observations by the High-Altitude Water Cherenkov (HAWC) observatory have revealed high-energy gamma-ray emission from the solar disk even during quiescent solar conditions \cite{Albert23}. These detections extend earlier Fermi-LAT measurements \cite{Abdo11,Ng16,Linden18}, which observed a hard gamma-ray spectrum between 0.1 and 200 GeV, up to TeV energies. Notably, the observed gamma-ray flux significantly exceeds theoretical predictions based on hadronic interactions between galactic CRs and the solar atmosphere \cite{Seckel91,Mazziotta20}. Furthermore, the spectral indices of the CR spectrum derived from Fermi-LAT and HAWC data differ substantially from that of the galactic CR background, raising questions about the canonical interpretation of solar disk gamma-ray emission.

One plausible interpretation of these findings is that the Sun may be capable of accelerating CRs to TeV energies even during periods of low solar activity \cite {Banik23}. This hypothesis, if validated, could have profound implications for our understanding of local CR sources and may help elucidate the origin of certain spectral features observed in CR data. Therefore, it is imperative to critically assess the feasibility of CR acceleration by a star like the Sun under quiescent conditions.

In this study, we investigate the potential of a normal star, exemplified by the Sun, to act as an accelerator of CRs to high energies. Specifically, we aim to identify plausible mechanisms by which a quiescent star could accelerate particles to relativistic energies. We demonstrate that charged particles can be accelerated to TeV energies in the Sun, or an analogous star, through shock acceleration occurring in the chromosphere. We also quantify the maximum achievable energy and spectral index of energy spectrum. We also estimate the corresponding flux of cosmic rays that such a star could produce.

\section{Environment for particle acceleration in stellar atmosphere}\label{sec2}
The stellar atmosphere of a sun like star is usually composed of several layers, including the photosphere and chromosphere. 
The solar chromosphere, a thin layer of plasma above the solar photosphere that extends around $2000 - 5000$ km \cite{Zhang98}, is a highly dynamic zone that is continually agitated by propagating waves and shocks. The temperature of the chromosphere increases from $6000^o$ C at the base to around $20000^o$ C at the top. The solar atmosphere contains shock waves \citep{Woger06}, which are believed to be a major factor in heating its outer layers \citep{Mathur22}. In the chromosphere and transition zones of the Sun, for example, the H \& K resonant lines in singly ionized calcium (Ca II) exhibit ``grains"$-$short-lived (100 seconds or less), jet-like patterns of heightened brightness in narrow band pictures.  With a typical size of $1''.95$ (1425 km), these characteristics show acoustic-shock-like disturbances (magneto-sonic waves) moving upward through the solar atmosphere \citep{Woger06}. The quiet Sun was thought to be mostly non-magnetic for a long time since continuum images only showed the photospheric granulation \cite{Rubio19}. In recent studies, however, the quiet Sun displays a reticular pattern of strong kilogauss (kG) fields in polarized light, which is called the magnetic network, together with many weaker, small-scale flux concentrations in the gaps between them, which is called the solar internetwork (IN) \cite{Rubio19}. The network defines the boundaries of supergranular cells (often 30,000 km in size, the greatest convective pattern on the solar surface) where horizontal flows become downdrafts \cite{Rubio19}. The internetwork (IN) closely resembles the inside of the supergranular cells \cite{Rubio19}. Bueno et al. (2004) \cite{Bueno04} revealed that the spatially unresolved hidden turbulent flux between the kG flux tubes and the hidden field in the seemingly empty regions is strong ($\sim 130$ G), but the field polarities are mixed. Current findings clearly supports the existence of such a strong magnetic field of the order of 100 G in the quiet Sun \cite{Rubio19,Orozco12,Aleman18,Sedik23}. The magnetic field arrangement in the quiet sun is canopy-shaped \cite{Li22}. Thus, the quiet solar chromosphere is a plausible location for proton acceleration to high energy via DSA and shock drift acceleration (SDA) because of the existence of shocks \citep{Woger06} as well as the strong magnetic field configuration.

\section{DSA and SDA in Solar atmosphere}\label{sec3}
It is often accepted that CRs can be accelerated to high energies in astrophysical environments by DSA, a first-order Fermi acceleration process \cite{Drury83,Bell78,Malkov01}. Numerous studies have been conducted in the literature on the diffusive shock acceleration (DSA) process for a parallel shock, in which the shock normal is parallel to the upstream magnetic field \cite{Bell78,Drury83,Blandford78}. In DSA, particles are diffusely scattering off magnetic fluctuations in both upstream and downstream regions and gain their energy by crossing the shock front multiple times. The mean fractional increase in energy in DSA is order of $\frac{\Delta E}{E} \sim \frac{U_s}{c}$ per one cycle of diffusion from upstream to downstream and back in this mechanism, where $U_s$ is shock velocity and $c$ is the velocity of light \cite{Bell13}. DSA predicts that the overall CR distribution function will be a power-law function in energy as $f (E) \propto E^{-\alpha}$, where 
\begin{eqnarray}
\alpha = 1 - \frac{\ln(1-P_{esc})}{\ln(1+\xi)}
\end{eqnarray}
is the spectral index. Here, $\xi$ and $P_{esc}$ represent the mean fractional energy increase and escape probability of particles per each cycle respectively \cite{Naito95}. In parallel shock acceleration scenario under DSA, the spectral index can be written as $\alpha = \frac{r+2}{r-1}$ where $r$ is the shock compression ratio \cite{Marcowith20}. For strong shock ($r = 4$), the spectral index of accelerated CRs becomes $\alpha = 2$.

In the quiet chromosphere, both shock and sound wave speeds are lower than those in a SNR due to the higher matter density in the solar atmosphere. However, the shock can still be quite strong, particularly in the ``Strong Pulse" model (see, for example, Figure 1 in Ref.~\cite{Sterling00}). These shock waves can compress the surrounding plasma, and the estimated compression ratio further confirms the potential strength of the shock \cite{Wentzel67}. For strong shocks in the chromosphere, $r \sim 4$, resulting in $\alpha \sim 2$ which is consistent with our result. However, for a typical shock velocity $U_s \sim 83$ km/s in quiet solar chromosphere \cite{Eklund21} (Note that for historical SNR, $U_s \sim 5000$ km/s, and magnetic field $B\sim 3 \mu G$ \cite{Bell13}), suggests the acceleration process may be slow in DSA for a parallel shock since $U_s/c \sim 3\times 10^{-4}$. 

Because the magnetic field arrangement in the quiet sun is canopy-shaped, quasi-perpendicular shocks appear to be a likely scenario. Consider a sub-luminal shock front ($U_s/cos\phi_{1} < c$) traveling at a speed of $U_s$ into a plasma at rest, where $\phi_{1}$ is the angle between the upstream magnetic field and the shock normal. An upstream particle incident on the shock front can either be transmitted into the downstream region or reflected from it by the mirror force of a compressed magnetic field \cite{Jokipii87,Xu22}. In such cases, shock drift acceleration (SDA) may also occur \cite{Jokipii87,Xu22}. In SDA, particles drift along the shock front due to a shift in the magnetic field and are accelerated by a convective electric field parallel to their motion. In the presence of magnetic fluctuations, both SDA and DSA simultaneously can help to accelerate particles at an quasi-perpendicular shock \cite{Jokipii87}. It provides more rapid acceleration than the parallel or quasi-parallel shock. 

The fluid speed in the upstream and downstream plasma are given by (subscripts i = 1, 2 indicate to the upstream and downstream regions) \cite{Naito95}
\begin{eqnarray}
u_1 = \frac{U_s}{\cos\phi_{1}}, 
\end{eqnarray}
and 
\begin{eqnarray}
u_2 = \frac{U_s}{r \cos\phi_{1}}\sqrt{1 + (r^2 - 1)\Gamma_s^2 \sin^2\phi_{1}} 
\end{eqnarray}
respectively, where $\Gamma_s = 1/\sqrt{1-U_s^2/c^2}$. The ratio of magnetic field strength of downstream to upstream can be defined as \cite{Naito95}
\begin{eqnarray}
b = \frac{B_2}{B_1} = \sqrt{1 + (r^2 - 1)\Gamma_s^2 \sin^2\phi_{1}}. 
\end{eqnarray}
Since the magnetic field in the downstream is stronger than that of the upstream for oblique shocks, $B_2 > B_1$, particles with a pitch angle smaller than $\mu_0$ in the upstream are reflected. The critical value of the pitch angle cosine can be written as \cite{Naito95}
\begin{eqnarray}
\mu_0 = \sqrt{1 - \frac{1}{b}}.
\end{eqnarray}

\begin{figure}[h]
  \begin{center}
  \includegraphics[width = 0.49\textwidth,height = 0.4\textwidth,angle=0]{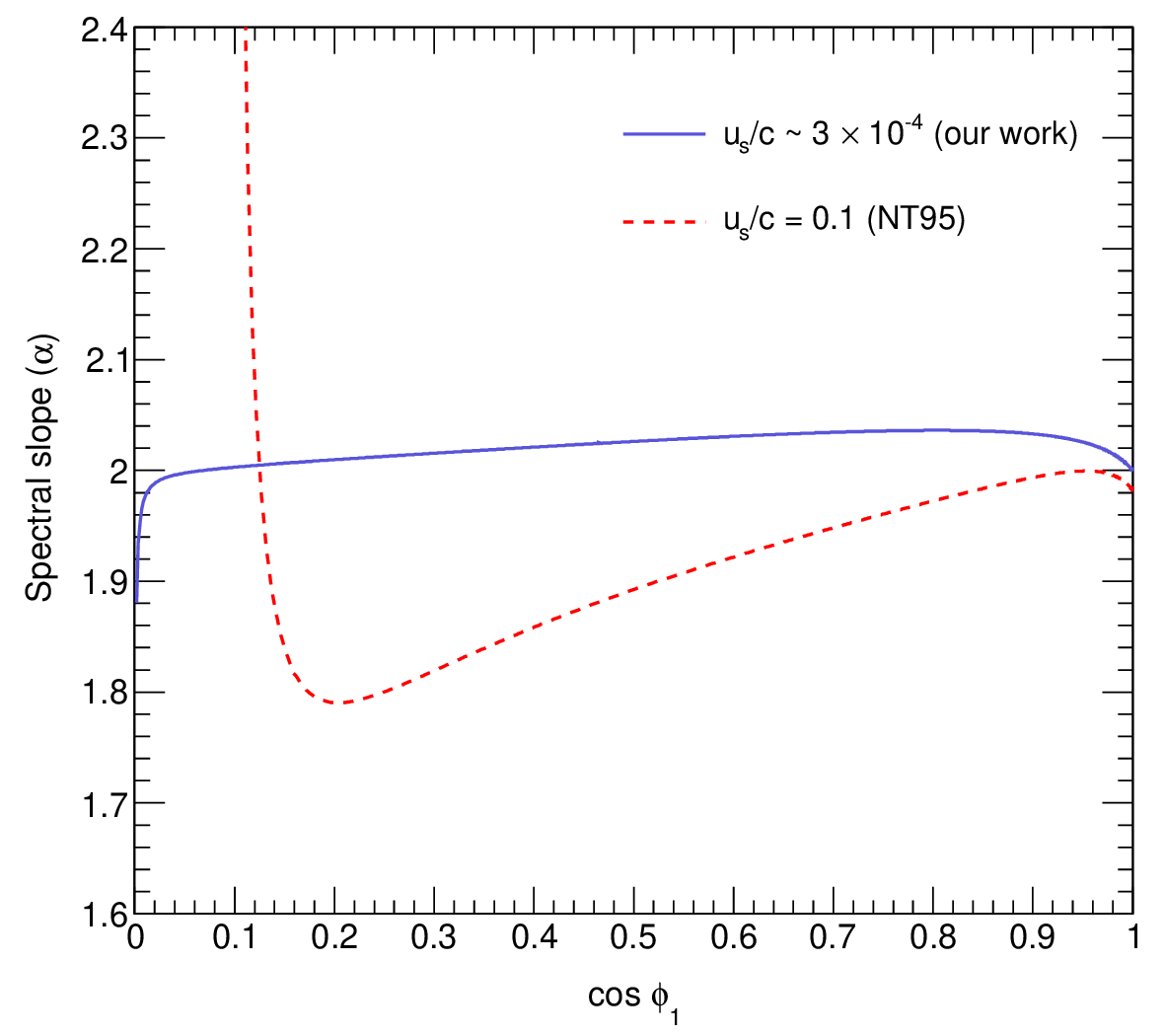}
\end{center}
  \caption{Spectral index of accelerated particles as a function of $cos \phi_1$ for $r = 4$. }
\label{Fig:5}
\end{figure}

Reflection probability of accelerating protons can obtained as \cite{Naito95}
\begin{eqnarray}
P_{ref} = \frac{\mu_0^2 (1 - \beta_1)^2}{(1 - \beta_1\mu_0)^2} 
\end{eqnarray}
where $\beta_i$ is the fluid speed normalized by the speed of light. The transmission probability of a particle coming from the upstream is \cite{Naito95}
\begin{eqnarray}
P_{tr} = \frac{1 - \mu_0^2 + 2\beta_1\mu_0(\mu_0 - 1)}{(1 - \beta_1\mu_0)^2} 
\end{eqnarray}
Then we can estimate energy gain per one cycle $\xi = \frac{\Delta E}{E}$ for obliquity angle $\phi_{1}$ by following Naito \& Takahara (1995) (NT95) (using Eq.~42 and associated relations) \cite{Naito95}. The escape probability for a particle from the upstream can be expressed by \cite{Naito95}
\begin{eqnarray}
P_{esc} = \frac{1 - \mu_0^2 + 2\beta_1\mu_0(\mu_0 - 1)}{(1 - \beta_1\mu_0)^2} \frac{4\beta_2}{(1+\beta_2)^2} 
\end{eqnarray}
We have estimated the spectral slope $\alpha$ of the accelerated protons as a function of obliquity angle $\phi_{1}$ using Eq.~1 and displayed in Figure~\ref{Fig:5}. Spectral slope at shock speed $0.1c$ matches well with the results presented in Xu \& Lazarian (2022) \cite{Xu22} and implies a steeper spectrum of accelerated particles at quasi-perpendicular shock. But the spectral slope for a shock wave in the solar chromosphere is found to be nearly $1.9$ to $2$ at a perpendicular or quasi-perpendicular shock, which is consistent with the results presented in Banik et al. (2023) \cite{Banik23}.

\begin{figure}[h]
  \begin{center}
  \includegraphics[width = 0.48\textwidth,height = 0.43\textwidth,angle=0]{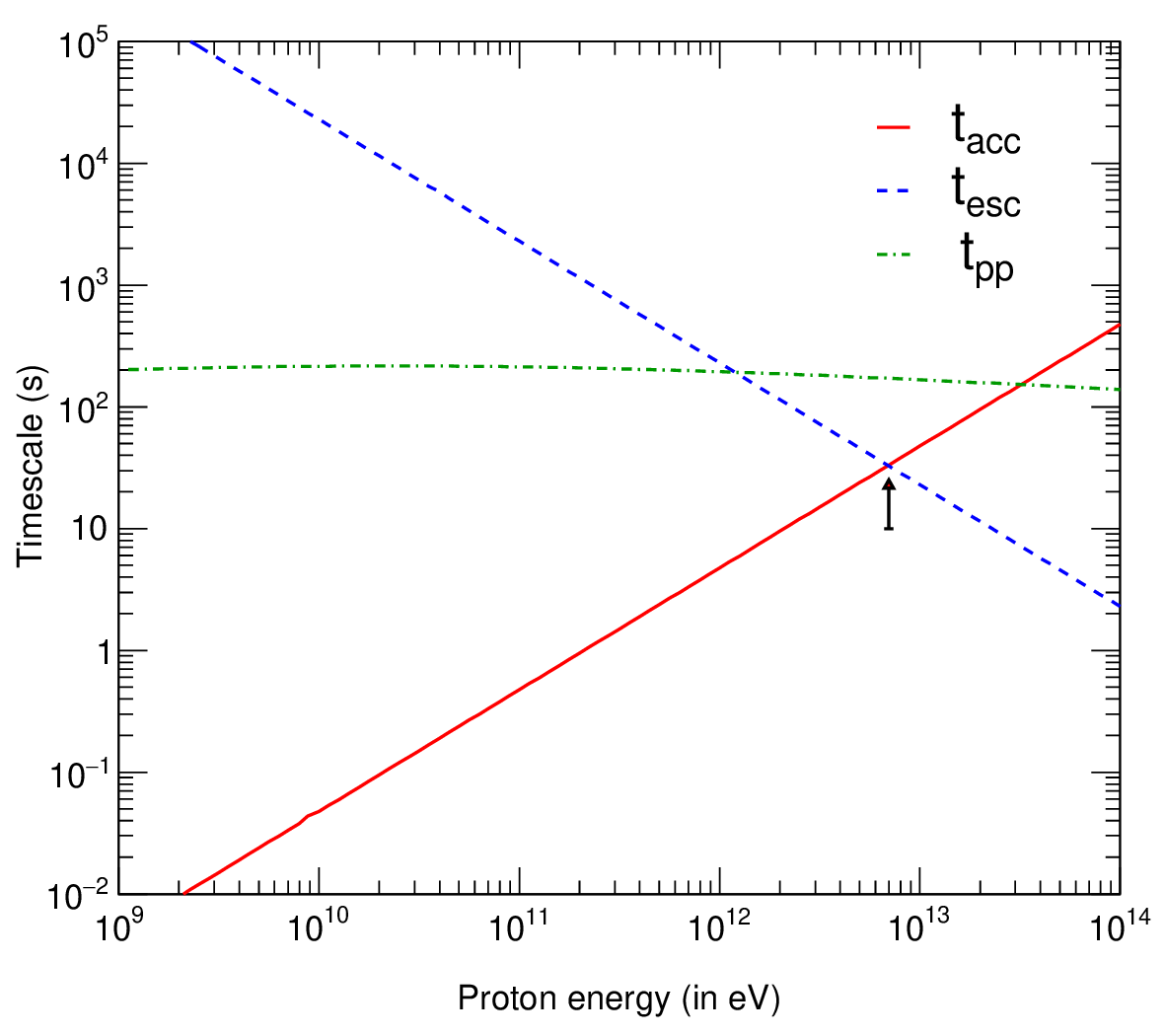}
\end{center}
  \caption{The estimated relevant timescales for protons. The black arrow represents the maximum achievable energy by a cosmic-ray proton in solar atmosphere.}
\label{Fig:6}
\end{figure}

Again the acceleration time can be defined as \cite{Drury83,Ostrowski88} 
\begin{eqnarray}
t_{acc} = \frac{\Delta t}{\xi}, 
\end{eqnarray}
where $\Delta t$ is the mean cycle time between successive interactions with the shock for upstream particle. For SDA and DSA operating simultaneously, 
\begin{eqnarray}
\Delta t = t_1 P_{ref} + (t_1 + t_2)P_{tr},
\end{eqnarray}
where $t_i = \frac{2\kappa_{n_i}}{v_{n_i}U_i}$ represents the mean residence time of the accelerating particle \cite{Xu22} and $\kappa_{n_i}$, $v_{n_i}$ and $U_i$ represent the diffusion coefficient, the mean particle velocity along the shock normal and the shock speed in the local fluid frame as defined in the Refs.~\cite{Xu22,Jokipii87}. Here, $\kappa_{n_i}$ can be expressed in terms of the parallel ($\kappa_{\parallel}$) and perpendicular ($\kappa_{\perp}$) diffusion coefficients as (Jokipii 1987) 
\begin{eqnarray}
\kappa_{n_i} = \kappa_{\parallel_i}\cos^2\phi_i + \kappa_{\perp_i}\sin^2\phi_i 
\end{eqnarray}
where the ratio of $\kappa_{\perp}$ to $\kappa_{\parallel}$ is given by $\frac{\kappa_{\perp}}{\kappa_{\parallel}} = \frac{1}{1+\eta^2}$. Here, $\eta = \frac{\lambda_{\parallel}}{r_g}$ where $\lambda_{\parallel}$ and $r_g$ are the mean free path parallel to the magnetic field and the gyroradius of the particles respectively. The expression of $v_{n_i}$ is given by \cite{Ostrowski88}
\begin{eqnarray}
v_{n_i} = v_{\parallel}\sqrt{\frac{\kappa_{n_i}}{\kappa_{\parallel_i}}} 
\end{eqnarray}
where $v_{\parallel}$ is the mean particle velocity along the magnetic field. Here, we consider $\eta = 500$ and $v_{\parallel} = \frac{c}{2}$ for isotropic particle distribution \cite{Xu22}. The considered parameter $\eta$ is consistent with its upper limit $v/u_s \sim 10^3$ (where $v$ is the particle velocity) \cite{Jokipii87}. The diffusion escape timescale from the downstream will be $t_{esc} = \frac{R_c^2}{\kappa_{n_1}}$ \cite{Bednarek07} considering $R_c \sim 3500$ km. The diffusive escape of accelerating particles limits the maximum energy of protons to be nearly $7$ TeV, considering a typical shock velocity $U_s \sim 83$ km/s in solar chromosphere. The relevant timescales of relativistic protons with energy for quasi-perpendicular case ($\phi_1 \approx 90^{\circ}$) are displayed in Fig.~\ref{Fig:6} to study maximum possible energy of accelerated protons in solar atmosphere.

In the solar interior, thermonuclear burning releases a tremendous quantity of energy, about $3.8 \times 10^{33}$ erg/s. A small portion of this energy heats the solar atmosphere. As the entire mechanical flux required to heat the solar atmosphere at the base of the chromosphere is around $10^{6}$ erg cm$^{-2}$ s$^{-1}$ \cite{Ulmschneider70}, the corresponding net power is roughly $10^{29}$ erg/s. Chromospheric shocks transport a fraction of this power, around $10^{26}$ erg/s (based on shock wave energy flux as provided in Ref.~\cite{Kuzma19}). Banik et al. (2023) \cite{Banik23} showed that a cosmic ray luminosity of about $10^{19}$ erg/s can explain the observed gamma-ray flux consistently. Thus, the chromospheric shocks have sufficient power to supply the required CR luminosity. The estimated cosmic ray flux reaching Earth from the quiet solar disk is found to be much lower than the overall cosmic ray flux observed at Earth \cite{Banik23}. It can also explain the observed CR deficit from the solar disk consistently \cite{Banik23}. Other similar normal stars are likely to emit cosmic rays with a flux comparable to that of the Sun, whereas massive stars are likely to emit cosmic rays with a higher flux. The efficiency of cosmic ray production in other stars may vary depending on their atmospheric properties. 

\section{Discussion and Conclusion}\label{sec5}

Our study indicates that the quiet solar atmosphere can accelerate CRs to a maximum energy of a few TeV for protons via DSA. Moreover, CR-induced instabilities may lead to magnetic field amplification \citep{Bell13, Lucek00}, both upstream and downstream of the shock, which in turn enhances the maximum achievable energy of CR particles. The spectral index of CRs accelerated by the quiet Sun is expected to be approximately $1.9$. However, the CR flux from the Sun, as observed at Earth, will be significantly lower than that of galactic CRs.

Stars exhibit a wide range of sizes and masses, which influence their atmospheric properties. Similarly, their atmospheric temperatures also vary greatly. Nonetheless, since the acceleration mechanism is DSA, the spectral characteristics of CRs produced in different stars should remain broadly similar. The maximum achievable energy and luminosity, however, will depend on specific stellar parameters such as size, magnetic field strength, and other atmospheric conditions. Assessing the net contribution of all normal stars to the galactic CR spectrum remains an essential task.

The Sun alone cannot account for the observed fine structure in the CR spectrum at TeV energies, but the collective contribution from nearby stars may play a role, which will be explored in a future work. In this study, we focus solely on the acceleration of protons. However, electrons may also be accelerated through the same mechanism, although likely to lower maximum energies due to higher energy losses. Further detailed investigations are required to explore this aspect.





\end{document}